\RequirePackage{fix-cm}
\documentclass[natbib, smallextended]{svjour3} 

\usepackage{amsmath,amssymb,amsfonts}
\usepackage{algorithmic}
\usepackage{graphicx}
\usepackage{textcomp}
\usepackage{multirow}
\usepackage{xcolor}
\usepackage[utf8]{inputenc}
\usepackage{lscape} 
\usepackage{color}
\usepackage{svg}
\usepackage{booktabs}
\usepackage{caption}
\usepackage{subcaption}
\usepackage{tikz}

\usetikzlibrary{shapes.geometric, arrows}

\tikzstyle{startstop} = [rectangle, rounded corners, minimum width=3cm, minimum height=1cm,text centered, draw=black, fill=red!30]
\tikzstyle{input} = [rectangle, rounded corners, minimum width=3cm, minimum height=1cm,text centered, draw=black, fill=blue!30]
\tikzstyle{embedding} = [rectangle, rounded corners, minimum width=3cm, minimum height=1cm,text centered, draw=black, fill=green!30]
\tikzstyle{lda} = [rectangle, rounded corners, minimum width=3cm, minimum height=1cm,text centered, draw=black, fill=orange!30]
\tikzstyle{classifier} = [rectangle, rounded corners, minimum width=3cm, minimum height=1cm,text centered, draw=black, fill=yellow!30]
\tikzstyle{arrow} = [thick,->,>=stealth]
\smartqed  

\usepackage[most]{tcolorbox}
\usepackage{lineno}

\newcommand*{\affaddr}[1]{#1} 
\newcommand*{\affmark}[1][*]{\textsuperscript{#1}}

\usepackage[colorlinks = true,
            linkcolor = blue,
            urlcolor  = blue,
            citecolor = blue,
            anchorcolor = blue]{hyperref}

\begin{document}

\title{Stuttering detection using speaker representations and self-supervised contextual embeddings}

\author{Shakeel A.~ Sheikh\affmark[1] \and Md Sahidullah\affmark[2] \and Fabrice Hirsch\affmark[3] \and Slim Ouni\affmark[1]}

\institute{
\at
\affaddr{\affmark[1]Universite De Lorraine, CNRS, Inria, LORIA, Nancy, France\\
          \email{\{shakeel-ahmad.sheikh, slim.ouni\}}@loria.fr}\\
\affaddr{\affmark[2]Institute for Advancing Intelligence (IAI), TCG CREST, India\\
          \email{sahidullahmd@gmail.com}}\\
\affaddr{\affmark[3]Universite Paul-Valery Montpellier CNRS, Praxiling, France\\
          \email{fabrice.hirsch@univ-montp3.fr}}\\
}

\maketitle

\begin{abstract}
The adoption of advanced deep learning architectures in stuttering detection (SD) tasks is challenging due to the limited size of the available datasets. To this end, this work introduces the application of speech embeddings extracted from pre-trained \textcolor{black}{deep learning} models trained on large audio datasets for different tasks. In particular, we explore audio representations obtained using \textcolor{black}{emphasized channel attention, propagation, and aggregation time delay neural network (ECAPA-TDNN)} and Wav2Vec2.0 models trained on VoxCeleb and LibriSpeech datasets respectively. After extracting the embeddings, we benchmark with several traditional classifiers, such as the \textcolor{black}{K-nearest neighbour (KNN)}, Gaussian naive Bayes, and neural network, for the SD tasks. In comparison to the standard SD systems trained only on the limited SEP-28k dataset, we obtain a relative improvement of 12.08\%, 28.71\%, 37.9\% in terms of \textcolor{black}{unweighted average recall (UAR)} over the baselines. Finally, we have shown that combining two embeddings and concatenating multiple layers of Wav2Vec2.0 can further improve the UAR by up to 2.60\% and 6.32\% respectively.
\end{abstract}


\section{Introduction}
Stuttering,~a neurodevelopmental speech disorder, caused by the failure of speech sensorimotor, is defined by the disturbance of uncontrolled utterances: \textcolor{black}{interjections (Insertion of sounds such as uh, uhm
), and \emph{core behaviors}: blocks (involuntary pauses), repetitions (involuntary recurring sounds, words or phrases), and prolongations (abnormal extension of a speech sound segments)}~\citep{smith2017stuttering, guitar2013stuttering, duffy2019motor, ward2017stuttering}. Studies show that persons who stutter (PWS) encounter several hardships in social and professional interactions~\citep{kehoe2006speech}.
In addition, more people are progressively interacting with voice assistants, but they ignore and fail to recognize stuttered speech~\citep{sheikh2021machine}, and the stuttering detection (SD) can be exploited to improve automatic speech recognition (ASR) for PWS to access voice assistants such as Alexa, Siri, etc.
\par

Usually, SD is addressed by various listening and brain scan tests~\citep{Ingham1996FunctionallesionIO, smith2017stuttering, sheikh2021machine}. However, this method of SD is high-priced and requires a demanding effort from speech therapists. The presence of uncontrolled utterances is reflected in the acoustic domain, which helps to discriminate them in various stuttering types. Based on the acoustic cues present in stuttered speech, several people employed a machine learning paradigm for SD. Some of the current state-of-the-art stuttering detection deep learning modelling techniques include:~ResNet+BiLSTM~\citep{tedd, melanie}, FluentNet~\citep{fluentnet}, StutterNet~\citep{stutternet, mcstutternet}. \citep{fluentnet, tedd} approached the SD as a multiple binary classification problem and trained separate ResNet+BiLSTM classifiers and FluentNet classifiers for each stuttering type. The models were trained using spectrogram input features on a small set of 24 UCLASS speakers. \citep{stutternet} approached SD via single branch StutterNet and proposed the time delay neural network based (TDNN) first multi-class classifier for SD and its types. The classifier is trained with 20 Mel-frequency cepstral coefficient (MFCCs)~\citep{huang2001spoken} input features on a large set of more than 100 UCLASS~\citep{howell2009university} speakers. \citep{sep28k} recently introduced a new SEP-28k stuttering dataset. \citep{melanie} exploited phoneme features and proposed BiLSTM method for SD. The method is trained on mel-spectral and phoneme-based input features by mixing SEP-28k, UCLASS, and FluencyBank datasets. In order to show the  efficacy of speech representations in SD, \citep{bayerw2v2} employed SVM as a downstream binary classifier  on the FluencyBank (English) and KSoF (German) datasets.

They reported an average F1 score of 53.8\% and 46.67\% on the KSoF and FluencyBank datasets, respectively. Although the method shows promising results, however, they formulated SD as a binary class classification problem (one vs rest) which requires separate training for each disfluency pair. Additionally, the fluent speech of PWS was not considered in their experimental study which could have a substantial impact on SD, as also demonstrated by \citep{stutternetmtl}. 
Comprehensive detail on SD methods can be found in~\citep{phdthesis} and in the review papers by~\citep{sheikh2021machine,sysreview}. 
\par 
Over the years, several stuttering datasets including: SEP-28k~\citep{sep28k}, UCLASS~\citep{sheikh2021machine}, LibriStutter~\citep{fluentnet}, and FluencyBank~\citep{sep28k} have been developed for investigating different SD models. Even though the SD methods discussed show promising results on these datasets, however,~these datasets are relatively very small and are limited to only a certain number of speakers. Due to the varying nature of stuttering from person to person, these small datasets incline to be biased towards these small pool of speakers~\citep{guitar2013stuttering}. In addition, the DL has shown tremendous improvement in ASR~\citep{speechrecog}, emotion detection~\citep{edwav2vec2,speechemotion}, speaker verification~\citep{haqiadversarial}, etc, however, the improvement in SD is bounded, most likely due to the limited size of stuttering datasets, which are unable to capture different speaking styles, accents, linguistic content, etc. In addition, collecting medical data requires big-budget and is very taxing, and stuttering data collection is no exception. 
\par
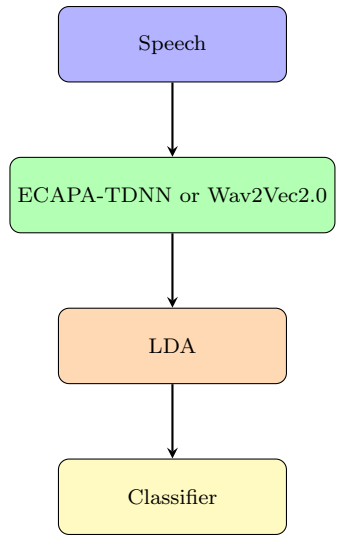
\begin{figure}
    \centering
    \begin{tikzpicture}[node distance=2cm]

\node (input) [input] {Speech};
\node (embedding) [embedding, below of=input] {ECAPA-TDNN or Wav2Vec2.0};
\node (lda) [lda, below of=embedding] {LDA};
\node (classifier) [classifier, below of=lda] {Classifier};

\draw [arrow] (input) -- (embedding);
\draw [arrow] (embedding) -- (lda);
\draw [arrow] (lda) -- (classifier);

\end{tikzpicture}
    \caption{\textcolor{black}{Block diagram with raw speech as an input followed by embedding extractors, LDA for dimensionality reduction followed by downstream classifiers.}}
    \label{fig:bd}
\end{figure}

 
To tackle this, we use the pre-trained speech embeddings that have been successfully used in ASR~\citep{wav2vec2, mohamed2022self} and emotion recognition~\citep{edwav2vec2}, for instance. In this paper, we mainly focus on the speaker and Wave2Vec2.0 contextual embeddings. Pre-training a model on such massive datasets can successfully capture the variable speaking styles and emotional behaviors which are extremely important in the SD~\citep{sheikh2021machine}. The general framework of the proposed system is shown in Fig.~\ref{fig:bd}.
\par
Our primary contributions to this paper are:
\begin{itemize} 
  \setlength\itemsep{0.05em}
    \item   We explore the use of speaker embeddings extracted from emphasized channel attention, propagation and aggregation (ECAPA)-TDNN~\citep{ecappatdnn}, and Wav2Vec2.0 speech representations for SD.
  
    \item We provide a novel way for SD, which exploits the information from the fully connected (FC) layer of ECAPA-TDNN~\citep{ecappatdnn} and draws on speech information from several hidden layers of the Wav2vec2.0 model~\citep{wav2vec2}.
    \item We also provide an analysis of the impact of using different layers from Wav2Vec2.0 and their concatenation in SD and also investigate the impact of combining information from ECAPA-TDNN and Wav2Vec2.0 embeddings via score fusion.
\end{itemize}

\section{Speaker and Contextual Embeddings}

\subsection{Speaker Embeddings}
\begin{figure}
    \centering
    \includegraphics[scale=0.7]{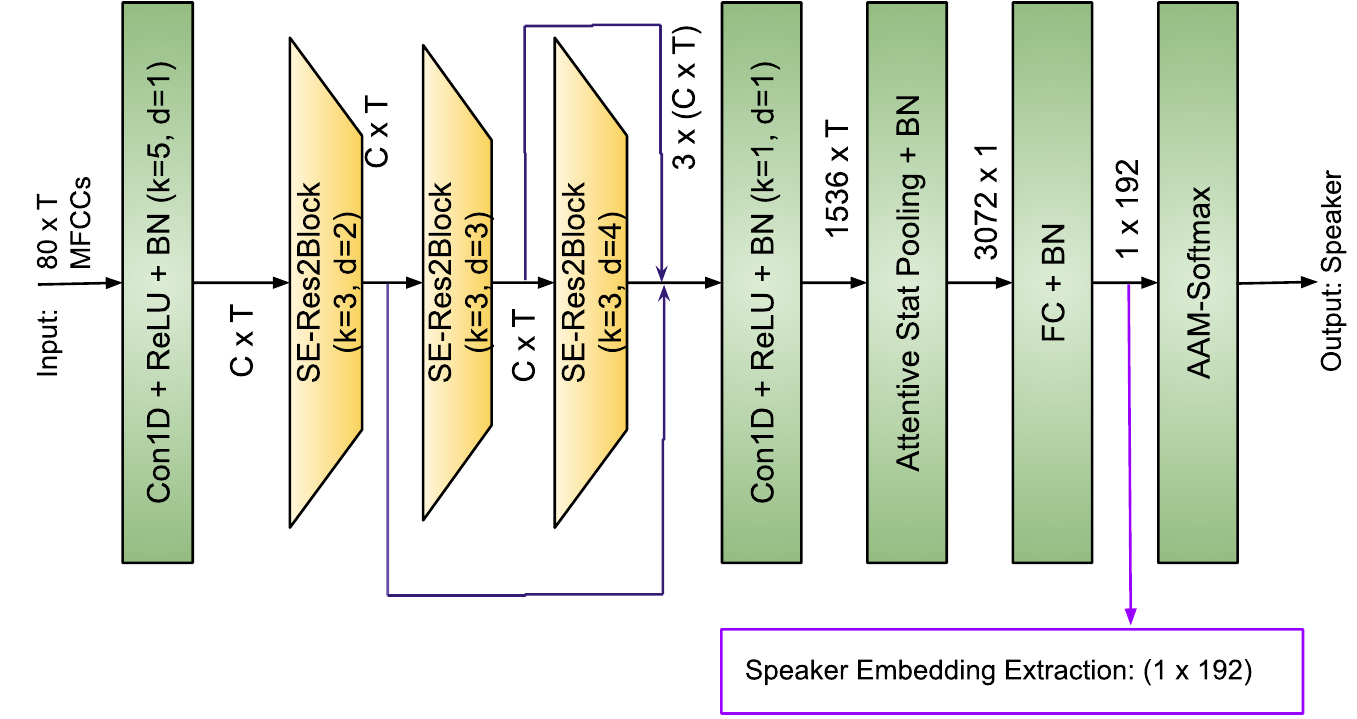} 
    \caption{ ECAPA-TDNN block diagram, k is kernel size, d is dilation, C is channel and T is temporal dimension.} 
    \label{fig:ecapatdnn} 
\end{figure}

Speaker embeddings are usually computed from trained neural networks to identify and classify speakers from a group of speakers~\citep{ecappatdnn}. The pre-trained speaker embeddings have been successfully applied in speaker diarization~\citep{ectdnnspkrdiar}. The ECAPA-TDNN is the SOTA for extracting speaker embeddings. As depicted in Fig.~\ref{fig:ecapatdnn}, the ECAPA-TDNN is composed of 1D convolution followed by three 1D squeeze and excitation (SE) Res2Blocks, 1D convolution, attentive statistical pooling, and an FC layer. A non-linear ReLU activation and batch normalization (BN) is applied after each layer in SE-Res2Block. The model is fed with 80 dimensional mean normalized log Mel-filterbank energy features and is trained on a large Voxceleb dataset with AAM-softmax loss function using Adam optimizer having a cycling learning rate between 1e-8 and 1e-3. In this paper, we use ECAPA-TDNN as a feature vector on SEP-28k to exploit it for SD. We extract $\mathbb{R}^{1\times192}$ speaker embedding feature vector from the FC layer of ECAPA-TDNN as shown by the purple block in Fig.~\ref{fig:ecapatdnn}. 
\subsection{Wav2Vec2.0 Contextual Embeddings}
The WaveVec2.0 model~\citep{wav2vec2} is a self-supervised representation learning framework of raw audio, and is comprised of three modules including feature encoder $f\colon \mathbb{X}\mapsto \mathbb{Z}$, contextual block transformer $g\colon \mathbb{Z}\mapsto \mathbb{C}$ and quantization block $\mathbb{Z}\mapsto \mathbb{Q}$ as depicted in Fig.~\ref{fig:wa2vec2.0}. The feature encoder is comprised of multi-layer 1D convolution blocks followed by BN and GELU activation functions, takes the normalized raw input $\mathbb{X}$ and encodes it into local feature representations~$\mathbb{Z} = f(\mathbb{X})$. These encoded feature representations of size $\mathbb{Z}^{T\times 768}$ are then fed to contextual transformer block  to learn contextual speech representations $\mathbb{C} = g(\mathbb{Z})$. The paper uses two different transformer 
networks with the base model consisting of 12 blocks having eight attention heads at each block, and the large model is comprised of 24 blocks with 16 attention heads at each block. The feature representations $\mathbb{Z}$ are also fed to the quantization module which is comprised of two codebooks having 320 possible entries in each. For each vector representation $\textit{z}_i$ $\in \mathbb{Z}$, a logit of $\mathbb{R}^{2\times320}$ is chosen using~(\ref{gumbel}) by concatenating the corresponding entries from each codebook, which is then followed by linear transformation to produce the quantized vector $\textit{q}_i$ of the local feature encoder representation $\textit{z}_i$ $\in \mathbb{Z}$. 
\begin{equation}
    p_{g,v} = \frac{\exp(l_{g,v} + \eta_v)/\tau}{\sum_{k=1}^{V}\exp(l_{g,v} + \eta_v)/\tau}
    \label{gumbel}
\end{equation}
where $l$ is logit, $v$ is v-th codebook entry, $g$ is codebook group, $\eta=-\log(-log(u))$ with $u$ are uniform samples from $\mathbb{U}(0,1)$, and $\tau$ is the temperature which controls the randomness. 
\par 
Similar to the masked language modelling,~the model is pre-trained in a self-supervised fashion using eq.~(\ref{eq:contrastiveloss}) by randomly masking certain time stamp representation vectors of the feature encoder and the training objective is to reproduce the quantized $\tilde{q_t}$ latent speech representation from a set of $K$+1 distractors including candidate vector $q_t$ and $K$ distractors $\in$ Q for masked time stamp vectors at the end of contextual transformer block. The distractors are uniformly sampled from masked frames of the same speech utterance. 
\begin{figure}
    \centering
    \caption{Block diagram of Wav2Vec2.0 architecture.}
    \label{fig:wa2vec2.0} 
\end{figure}

\begin{equation}
    \mathbb{L}_{cont} = -\log\frac{\exp(sim(c_t, q_t)/\tau)}{\sum_{\tilde{q}\in Q}\exp(sim(c_t, \tilde{q})/\tau)}
    \label{eq:contrastiveloss}
\end{equation}
where $sim(c_t, q_t)$ computes the cosine similarity between the quantized vector $q_t$ and contextualized transformer vector $c_t$.
\par 
The authors have released several pre-trained feature embeddings with dimensions of $768$ (base) and $1024$ (large) and we are using the base one ($768$-dimensional) pre-trained on 960 hours of LibriSpeech dataset and then fine-tuned for ASR using CTC loss function by adding a linear layer on top of the contextual block. 
\par 
The Wav2Vec2.0 showed remarkable improvement in ASR~\citep{wav2vec2}, emotion detection~\citep{edwav2vec2}. In the stuttering speech, most parts of the speech utterance are perturbed, and it seems a plausible way to employ and explore the role of the contextual and encoder representations in SD. In this work, without fine-tuning, we employ a total of 13 contextual embeddings extracted from a local encoder and 12 layers of the contextual transformer block as depicted in Fig.~\ref{fig:wa2vec2.0}.

\section{Classifier Description}

\subsection{K-Nearest Neighbourhood}
A non-parametric supervised algorithm based on distance metric is used mostly for classification tasks. The prediction of the query sample depends on the voting majority of its $K$ nearest neighbors~\citep{murphy2012machine}. In this work, use the Minkowski metric distance from~eq.~(\ref{eq:minsk}) with $p=2$ (Euclidean) to fit the $K$-NN on the SEP-28k dataset using embeddings computed from pre-trained ECAPA-TDNN and Wav2Vec2.0.
\begin{equation}
    D = \left[ \sum_{i=1}^{k}\|x_i - y_i\|^p)\right]^{1/p}
    \label{eq:minsk}
\end{equation}
\subsection{Gaussian Back-End}
 
A naive Bayes classifier (NBC) is simply a Bayesian network to handle continuous features by representing the likelihood of features using Gaussian distribution~\citep{murphy2012machine}. Given a data set $\mathbb{D} = (X_i,d_i)$ of $N$ samples with $\mathbb{R}^{1\times K}$\footnote{K is 768 for Wav2Vec2.0 and 192 for ECAPA-TDNN} pre-trained features ,~NBC assumes that the likelihood of class conditional densities is normally distributed by
\begin{equation}
    p(e|C=c, \mu_c,\Sigma_c) = \mathbb{N}( e | \mu_c, \Sigma_c)
\end{equation}
where $ e \in X$, is extracted pre-trained representation, $\mu$ and $\Sigma$ are class-specific mean vector and covariance matrix respectively. The posterior probability for each target class is computed then by Bayes' formula: 
\begin{equation}
    \underbrace{p(C=c| e, \mu_c,\Sigma_c)}_{\text{class posterior}} = \frac{\overbrace{p( e|C=c, \mu_c, \Sigma_c)}^{\text{class conditional likelihood}}p(C=c)}{\sum_{i=1}^{K} p( e|C=i, \mu_i, \Sigma_c)p(C=i)}
    \label{eq:gaussianbackend}
\end{equation}
and then the query sample $ e$ is classified by taking \emph{argmax} over the classes. 
\begin{equation}
    \hat{y}( e) = \mathrm{argmax}~p(C=c |  e, \mu_c, \Sigma_c)
    \label{eq:argmax}  
\end{equation}

\subsection{Neural Network}
A neural network (NN) with just a few layers can also be applied on top of the pre-trained embeddings extracted from ECAPA-TDNN and Wav2Vec2.0. In this work, we use two-branched NN with \emph{FluentBranch} differentiating between fluent and stuttered utterances, and \emph{DisfluentBranch} classifying stuttered speech utterances into several disfluency types. Each branch is composed of three FC layers with each layer followed by ReLU activation~\citep{activationssurvey} and 1D BN~\citep{batchnorm} functions. A dropout of 0.2~\citep{JMLR:v15:srivastava14a} is applied for first two FC layers. A Softmax layer is used in the end to get the desired classes. Following the approach in~\citep{sep28k,stutternetmtl}, for \emph{FluentBranch}, we employ a pseudo labelling scheme, where we re-label all different disfluent speech samples as one class and train the binary \emph{FluentBranch} branch to differentiate between fluent and stuttered utterances. For the multi-class \emph{DisfluentBranch} branch, we train it by penalizing the fluent class with zero loss. During evaluation phase, the outcome from \emph{FluentBranch} is considered if the prediction is fluent class, otherwise predictions from \emph{DisfluentBranch} are taken into consideration.

\section{Experimental Setup}
\subsection{Embedding Extractors}
\textbf{ECAPA-TDNN}: For each three-second speech utterance of the SEP-28k dataset, we extract speaker embeddings of dimension $\mathbb{R}^{1\times192}$ after the FC layer of ECAPA-TDNN as shown by the purple line in Fig.~\ref{fig:ecapatdnn}, resulting in $\mathbb{R}^{N\times192}$ data (N is total samples). We use SpeechBrain toolkit~\citep{speechbrain} for the extraction of 192-dimensional speaker embeddings. In addition, we also use linear discriminant analysis (LDA) for dimensionality reduction with the component size of four resulting in $\mathbb{R}^{N\times4}$ dimensional data before passing it to the downstream classifiers. 
\newline
\textbf{Wav2Vec2.0}: For contextual embedding extraction of each SEP-28k sample, we extract $\mathbb{R}^{T\times 768}$ (T is a temporal dimension) dimensional representations from local feature encoder $\mathbb{Z}$ and from each layer of contextual transformer block $\mathbb{C}$, each of which yields different contextual representation as shown by purple block in Fig.~\ref{fig:wa2vec2.0}. Before passing each layer representation separately to the downstream classifiers, we apply statistical pooling across the temporal domain and concatenated the mean and standard deviation resulting in a feature vector of $\mathbb{R}^{1\times768\times 2}$. Moreover, we also concatenate the contextual embeddings of the local encoder (L1), L7, and layer L11 of $\mathbb{C}$ after applying the statistical pooling and LDA (with a component size of four) which results in $\mathbb{R}^{N\times12}$ feature vector. \textcolor{black}{LDA is useful when the number of features is large compared to the number of samples, and there is a need to reduce the dimensionality of the data while preserving the most relevant information for classification.
} We extract embeddings from the PyTorch version of Wav2Vec2.0~\citep{pytorch}.

\subsection{Dataset Description}

In this case study, we used SEP-28k stuttering dataset~\citep{sep28k}  which consists of 28,177 speech samples from 385 podcasts. After removing non-stuttered samples, we are left with 23573 annotated speech segments. We randomly selected 80\% podcasts (without mixing podcasts). The speaker information is missing from the SEP-28k dataset, so we divided the dataset based on podcast ids (assuming each podcast is having a unique speaker). This dataset contains two different types of annotations including stuttering and non-stuttering. We considered only stuttering annotations (repetitions, blocks, interjections, prolongations, and fluent speech) and avoided the non-stuttering annotations (unsure, no speech, poor audio quality, music, unintelligible and natural pauses) in this experimental study. The SEP-28k dataset is publicly available, but the details about the train, validation and test set splits are not provided. So, we decided to create a publicly available protocol that ensures no overlap of podcasts between train, validation and test sets. For splitting of the dataset, please see Table~\ref{tab:datastat}. The experimental results mentioned are the average of 10-fold cross validation experiments, and are compared to the baseline results from~\citep{stutternet, stutternetmtl, tedd} which is trained on MFCC features.

\begin{table}
 \caption{SEP-28k Dataset Split into Train, Val and Test Sets ( B: Block , F: Fluent , R:
Repetition , P: Prolongation , I: Interjection, ID: SetID)}
  \label{tab:datastat}
  \centering
   \begin{tabular}{ c c c c c c c}
     \multicolumn{1}{c}{\textbf{}} &                                     \multicolumn{5}{c}{\textbf{Distribution}} &                                     \multicolumn{1}{c}{\textbf{}} \\ 
    \midrule
    ID&R&P&B&I&F&Tot.Count\\
    \midrule
      &&&Train&&& \\
       \midrule
  1&  2681	& 1384& 	1726& 	3181	& 9950& 18922\\
     2&2696&	1495&	1758&	3274&	10356 &19579 \\
    3&2652&	1481& 1761&	3202&	10301&19397	\\
    4&2676&	1518 &1753&   3305 &   10293& 19545\\
    5&2513&	1430&	1679&	3115&	9832&18569\\
    6&2745&	1457	&1716	&3119&	9806 &18843\\
    7&2805&	1437&	1718&	3265&	10174&  19399\\
    8&2491&	1439&	1666&	3085&	9914  &  18595\\
    9&2409&	1412&	1701&	3179&	10091&  18792 \\
  10 &2556&	1348&	1680&	3082&	9505& 18171  \\
   \midrule
  &&&Val&&& \\
       \midrule
    1 & 331	& 277& 	245	& 486& 	1466& 2805 	\\
        2&314&	157	& 182&	349&	1097&2099	\\
    3&287&	133  &154&	349 &	973  &1896\\
  4&262&	125	&149	  & 368   &	1081&1985	\\
  5&363&	154	 &   230 & 	378 &	1241&2366\\
    6&241&	176&	160&	361&	1287&2225\\
    7&284&	163	&   192 &	414 &	1139&2192\\
  8&362&	168	 &   224&	    390	&    1200 &  2344 \\
  9&402&	141	 &   176&	    338&	    938	&  1995  \\
    10&396&	265 &	228 &	488	&    1478 &  2855\\
     \midrule
  &&&Test&&& \\
\midrule
    1& 274& 	109& 	132	& 328& 	1003& 1846\\
    2&276&	118	& 163&	372&	966& 1895 \\
    3&347&	156	 &188 &   444&	    1145& 2280\\
    4&348&	127	&201   &322	  &  1045&2043\\
    5&410&	186	  & 194   &	502	 &  1346&2638\\
    6&300&	137	&  227&	515	&1326&2505\\
    7&197&	170	&    193 &	316	 &   1106& 1982\\
    8&433&	163	  &  213&	    520	 &   1305 &   2634\\
    9&475&	215	  &  226 &	478& 	1390&   2784 \\
    10&334&	157 &	195	&    425&	    1436  &  2547 \\
    
    \bottomrule
  \end{tabular} 
\end{table}
\subsection{Implementation}
For implementing the proposed pipeline for NN, we use PyTorch library~\citep{pytorch}, and for LDA, KNN, and NBC, we have used the Scikit-learn~\citep{scikit-learn} toolkit. We have chosen a value of $K=5$ using the elbow method in the KNN classifier. The downstream NN is trained with a batch size of 128 using a normal sum of cross entropy loss ($L_{tot}$ = $L_f + L_d$,~$L_f$:FluentBranch loss, $L_d$:DisfluentBranch loss) function optimized by the Adam optimizer with a learning rate of 1e-2, and the training is stopped using an early stopping scheme on validation loss with the patience of seven. To evaluate the model performance, we use the following metrics: unweighted average recall (UAR) and accuracy which are the standard and are widely used in the stuttered speech domain~\citep{sheikh2021machine, Schuller22-TI2, sheikh_acmmm}. 

\section{Results and Discussion }


For thorough evaluation, we train each of the proposed models 10 times. Table~\ref{tab:fusion}. shows the average UAR and accuracy results of different stuttering and fluent classes on exploiting different features extracted from ECAPA-TDNN and Wav2Vec2.0. We compare the proposed speaker and contextual embeddings based SD models to the work ResNet+BiLSTM~\citep{tedd}, \emph{StutterNet}~\citep{stutternet} (For a fair comparison, we trained these on the SEP-28k dataset with MFCC input features) and to the multi-branched \emph{StutterNet}~\citep{stutternetmtl} having two different branches with one branch differentiating between fluent and stutter utterances and the other branch distinguishing among different disfluency types. We use LDA for dimensionality reduction of features in each case before passing them to the classifiers for prediction\footnote{\textcolor{black}{We have not compared from an execution time perspective. However, since the LDA transformed data has only four dimensions compared to the original 192 (for speaker embeddings) and 768 (for contextual embeddings), we expect that the “with LDA” configuration drastically reduces the computational time.}}. 
 \begin{table}[th]
  \caption{SD results (TA: Total accuracy, B:
Block , F: Fluent , R: Repetition , P: Prolongation , I: Interjection, UAR: Unweighted Average Recall, BLs: Baselines, MB: Multi Branch) for different methods. The results reported for Wav2Vec2.0 are from L11.}
  \label{tab:fusion}
  \centering
  
  \scalebox{0.98}{\begin{tabular}{ c c c c c c c c}
  \toprule
     \multicolumn{1}{c}{\textbf{Model}} &                                     \multicolumn{1}{c}{\textbf{R}} &                                     \multicolumn{1}{c}{\textbf{P}} &   
    \multicolumn{1}{c}{\textbf{B}} & 
    \multicolumn{1}{c}{\textbf{I}} & 
    \multicolumn{1}{c}{\textbf{F}}   &
    \multicolumn{1}{c}{\textbf{TA}} &
    \multicolumn{1}{c}{\textbf{UAR(\%)}} \\
    \midrule
     &\multicolumn{6}{c}{\textbf{MFCCs} (BLs)}  \\ 
    \midrule
   
     \citep{stutternet} & 21.99& 27.78 & 1.98 & 49.99 & 88.18 & 60.33 &48.00\\
      ~\citep{stutternetmtl}  & 28.70&	37.89	&	9.58	&	57.65	&	74.43	&	57.04&41.80\\
 \citep{tedd}&34.79&	30.19&	5.92&	49.26&	 	75.47&	55.62&	 	39.13 \\

         \midrule
           &\multicolumn{6}{c}{\textbf{Embedding: ECAPA-TDNN}}  \\ 
    \midrule
     KNN& 17.82&	6.40&	6.09&	26.58&		71.55&	45.73&25.80\\
    NBC & 22.15	&13.84&	9.42&	31.46&		63.57&	44.04&28.00\\
    NN & 23.06 &	11.67 &	5.79 &	43.29 &		69.48 &	49.12 &33.20  \\
    \midrule
    
     KNN~+~LDA  &21.92&	11.93&	8.56&	26.6& 	66.77&	44.37&27.40\\
    NBC~+~LDA  &11.99&	7.24&	2.52&	25.43&	88.77&	53.53&27.20\\
    NN~+~LDA  &24.51	&10.33&	5.03&	44.49&	68.73&	48.81&32.40\\
\midrule
     & \multicolumn{6}{c}{\textbf{Embedding: Wav2Vec2.0}}  \\     
        \midrule
         KNN  &  24.48&	8.88&	11.33&	54.02&	 84.10&	58.85&36.40 \\
    NBC &41.62 &	17.33 &	35.75 &	36.88 &	60.96 &	48.74&38.60\\

    NN &46.31&	35.55&	14.17&	66.07&	81.14&	64.69&52.20\\
 \midrule

    KNN~+~LDA &47.22	&38.79&	18.28&	65.87&		78.52&	63.84&49.80\\
    NBC~+~LDA &42.61	&38.17	&17.13	&67.29		&85.81	&67.29&50.20\\
NN~+~LDA&50.15&	39.25&	18.03&	71.83&	80.50&	66.59&53.80\\
\midrule
     \multicolumn{8}{c}{\textbf{Score fusion: ECAPA-TDNN and Wav2Vec2.0}} \\
     \midrule
        KNN~+~LDA  &44.06	&36.95	&16.13	&64.61	&82.70	&65.18&49.00 \\
       NBC~+~LDA  &41.75	&38.63	&16.60	&66.83	&86.78	&67.73&50.40 \\
       NN~+~LDA  &43.58	&38.94	&19.11	&67.36	&84.63	&67.26&50.80 \\
       \midrule
           \multicolumn{8}{c}{\textbf{Embedding fusion: ECAPA-TDNN and Wav2Vec2.0}}  \\ 
           \midrule
KNN~+~LDA &45.38		&37.29		&17.92		&62.18			&80.68		&64.08&48.60 \\
NBC~+~LDA &44.22		&40.01		&19.70		&67.61			&84.55		&67.44&51.40  \\
 NN~+~LDA &44.13		&38.02		&18.39		&66.44			&84.53		&67.00&55.20  \\
         \midrule
          \multicolumn{8}{c}{\textbf{Embedding fusion: L1 + L7 + L11}}  \\ 
          \midrule
          KNN~+~LDA &46.98	&42.18	&20.96	&66.28		&82.24	&66.30&51.60 \\
         NBC~+~LDA &48.46	&47.84	&28.51	&70.41	&80.33	&67.45&55.00 \\
         NN~+~LDA &46.79	&40.79	&23.86	&69.54	   &84.32	&\textbf{68.35}&\textbf{57.20} \\
       \bottomrule
   \end{tabular}}  
  \end{table}
  \begin{figure*}
    \centering
    \includegraphics[width=0.7\textwidth]{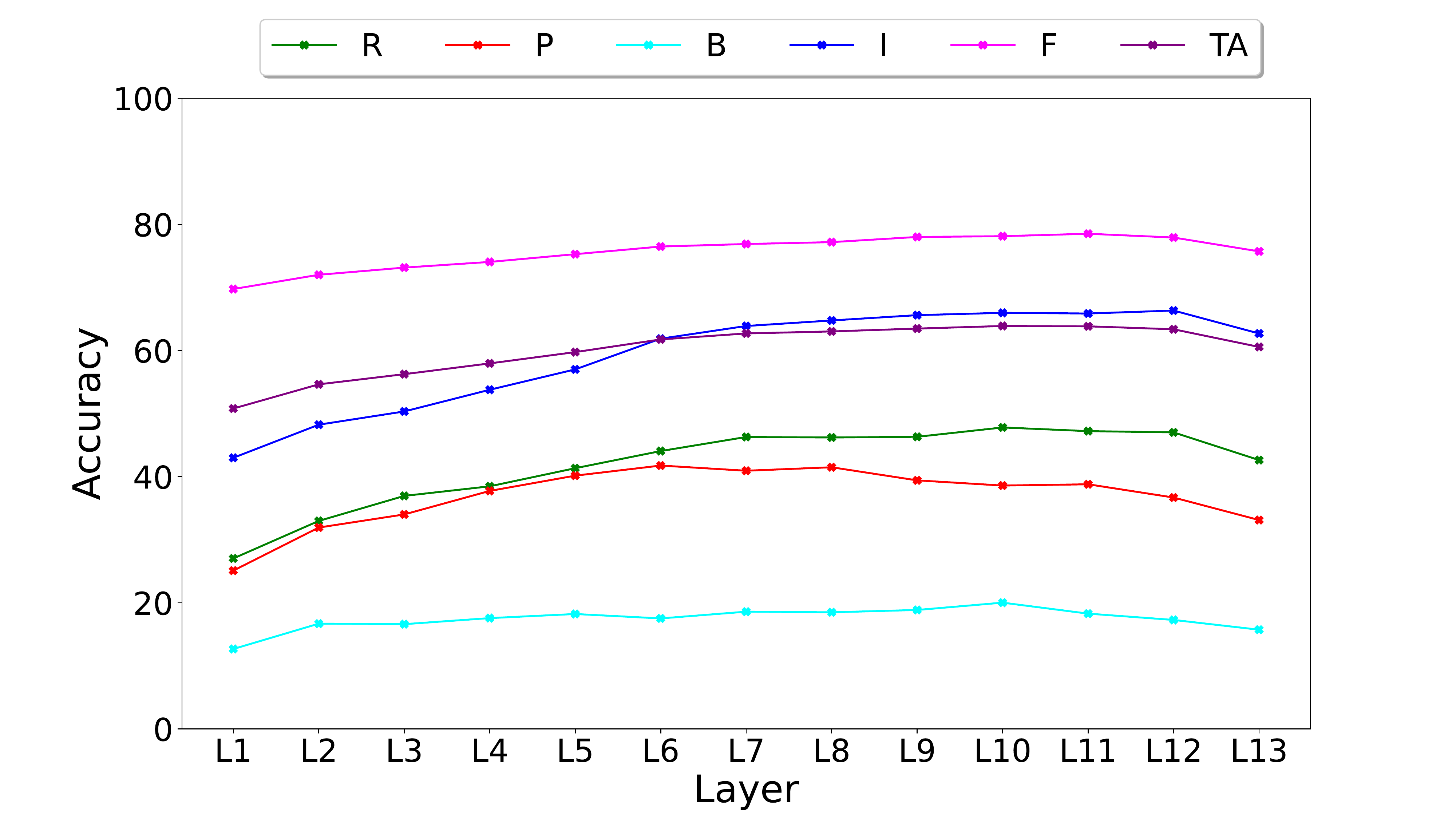}   \\
    \centering
    \includegraphics[width=0.7\textwidth]{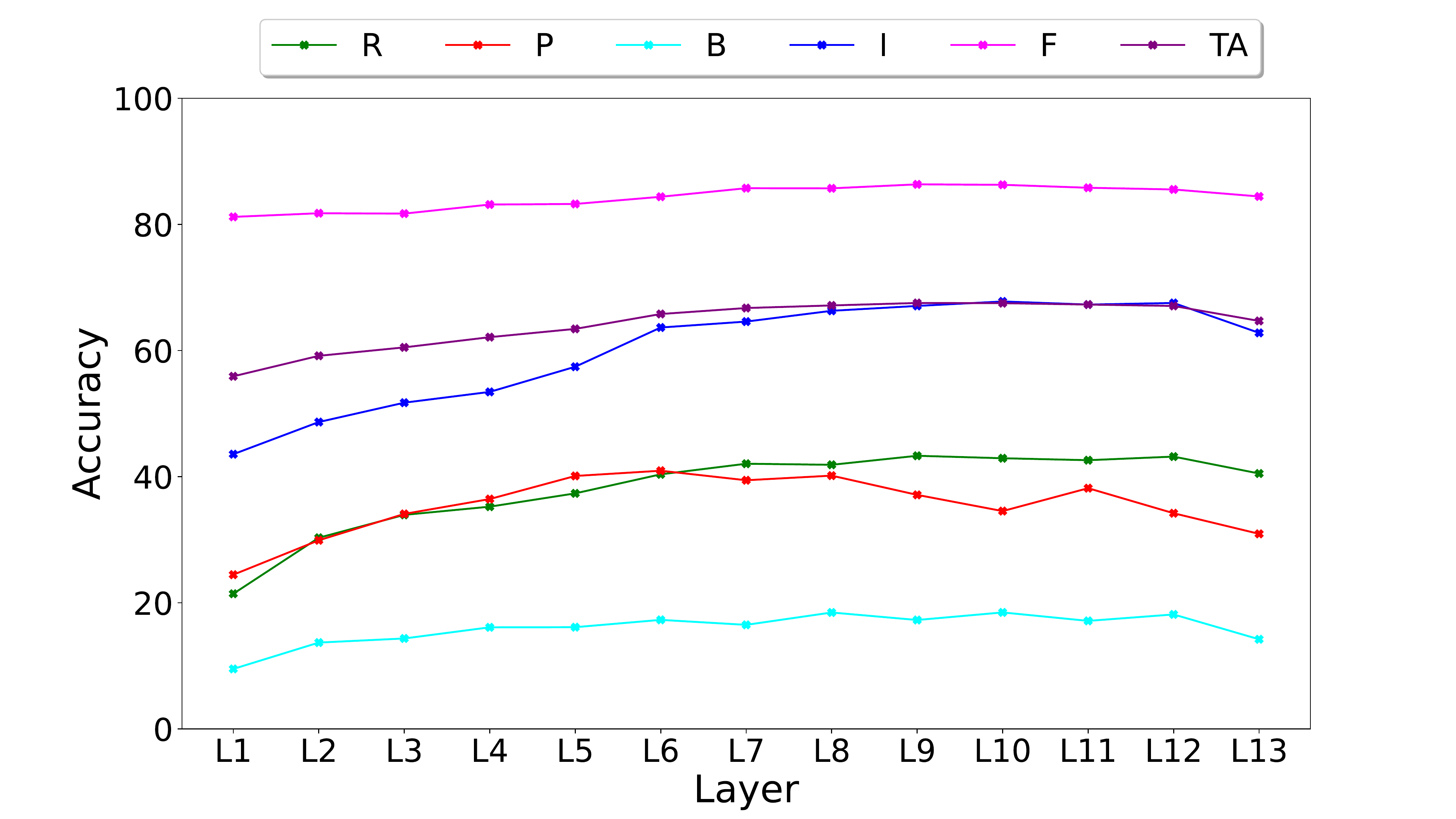} \\
    \centering
    \includegraphics[width=0.7\textwidth]{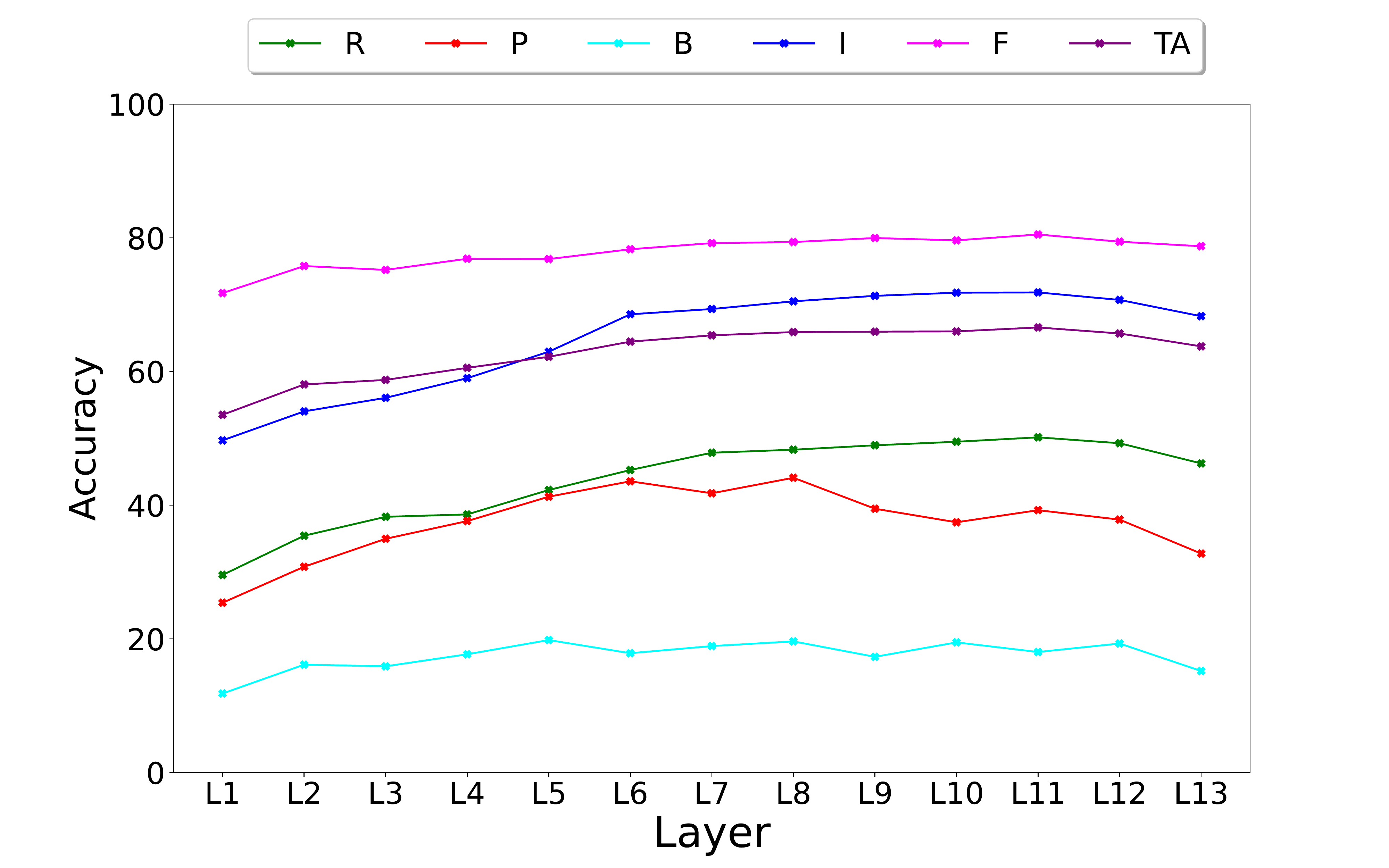}   \\
\caption{Impact of various Wav2Vec2.0 contextual layers in SD with KNN (Top), NBC (Middle), and NN (Bottom).}
\label{fig:layerimpact}
\end{figure*}

\begin{figure}
     \centering
     \begin{subfigure}[b]{0.7\textwidth}
         \centering
         \includegraphics[width=\textwidth]{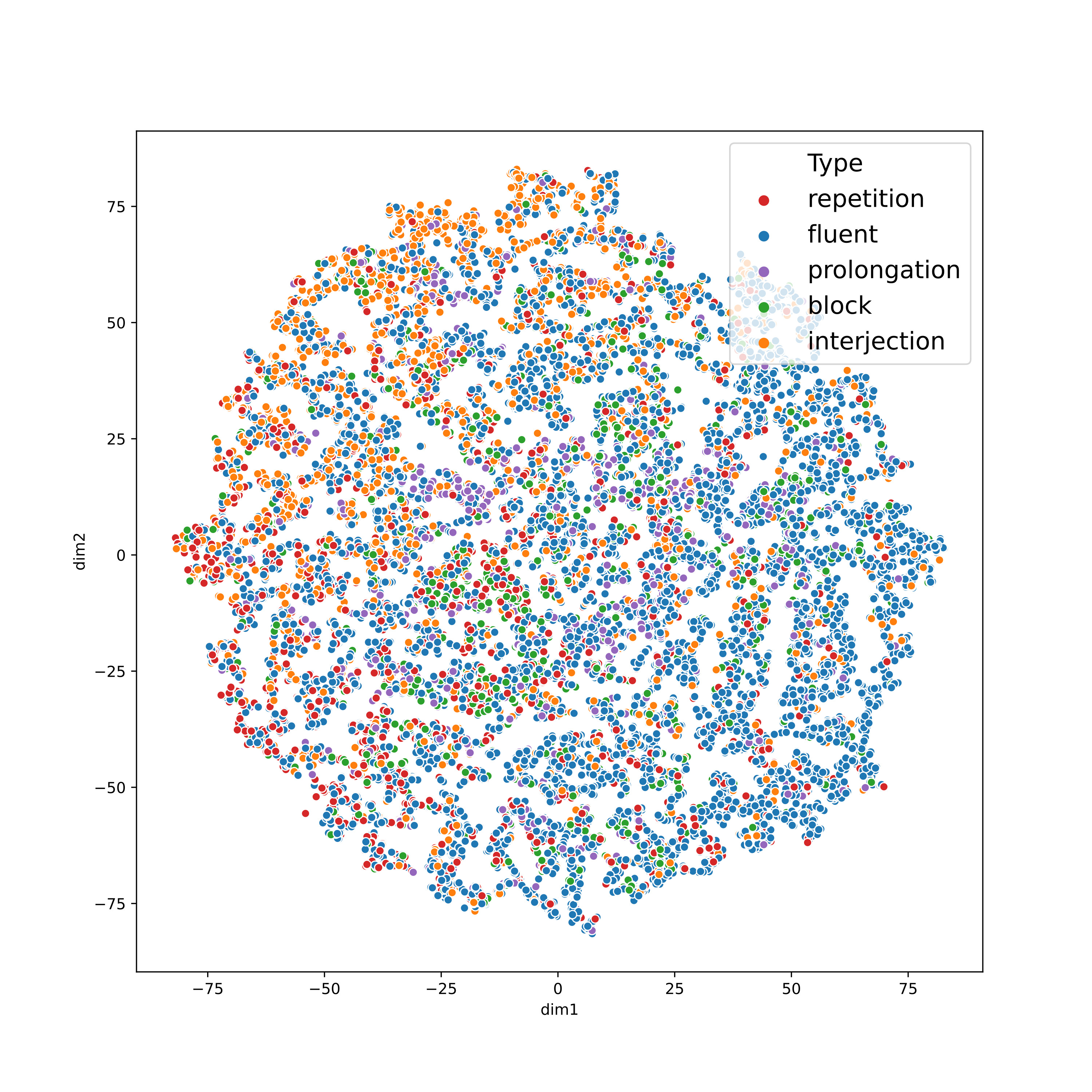}
      \end{subfigure}
     \hfill
     \begin{subfigure}[b]{0.7\textwidth}
         \centering
         \includegraphics[width=\textwidth]{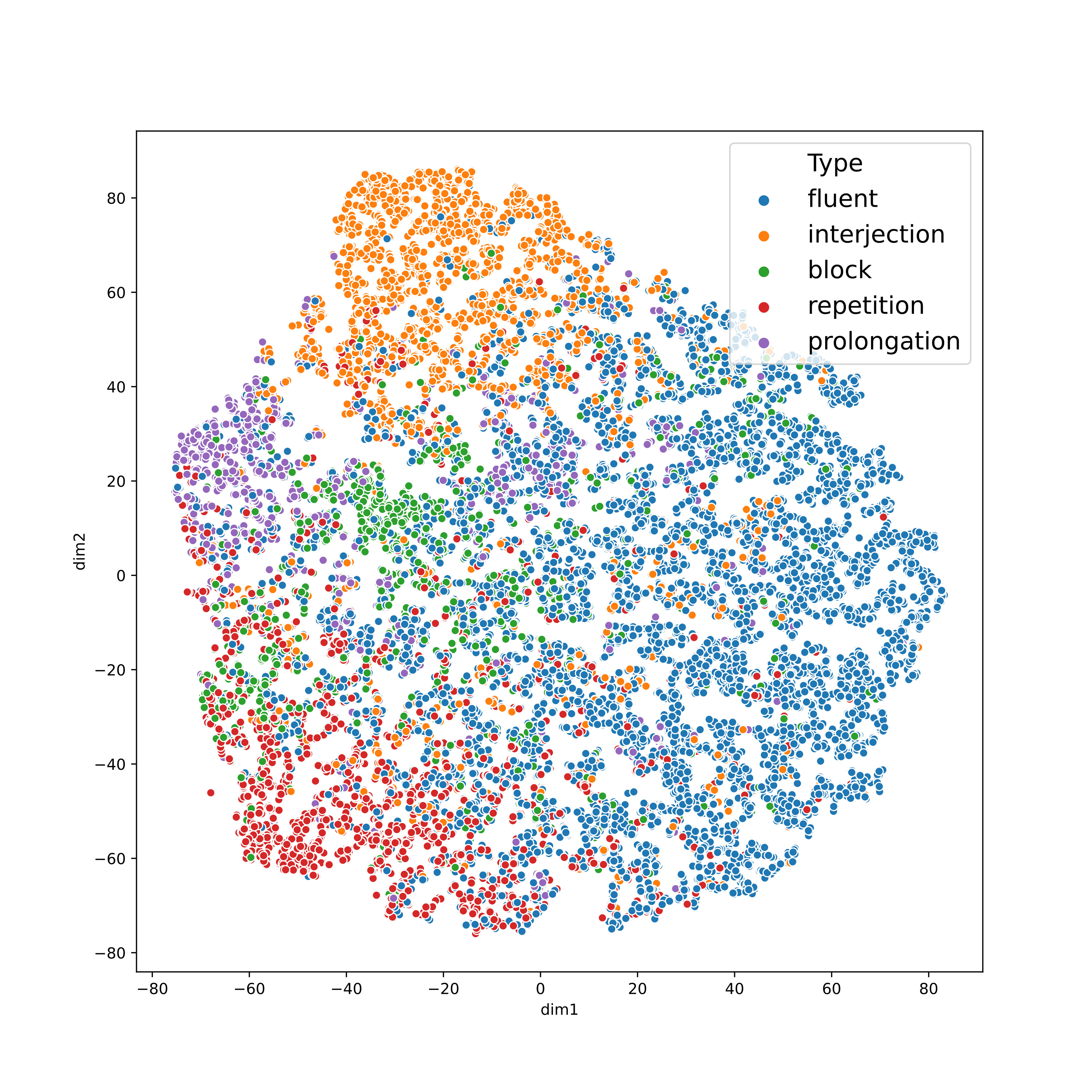}
      \end{subfigure}
       \caption{t-SNE embeddings from the FC layer of ECAPA-TDNN (T) and L11 of the contextual block $\mathbb{C}$ of Wav2Vec2.0 (B). The visualisation is only for exploration purposes to understand which embeddings are better for stuttering detection downstream classifiers~\citep{hinton_tsne}.} 
             \label{fig:embed} 
\end{figure}
\par 
\textbf{Speaker embeddings}:
We see from Table~\ref{tab:fusion}, that the downstream classifiers trained on ECAPA-TDNN embeddings perform poorly in all the stuttering classes as compared to the baseline results from Table~\ref{tab:fusion}. This is evident from Fig.~\ref{fig:embed} as well, where the different stuttering type utterances are mixed and no clear cluster is visible among the disfluency classes. Furthermore, applying magnitude normalization on ECAPA-TDNN embeddings before passing them to the downstream classifiers, improves the SD performance marginally in the majority of classes. The ECAPA-TDNN is trained and adapted for the speaker identification task and it is likely possible that the information (such as linguistic content, prosody, and emotion state) which isn't essential for that task but could be crucial for SD gets removed from the latent embeddings. 
\par 
\textbf{Wav2Vec2.0 contextual embeddings}:
Table~\ref{tab:fusion}. shows the results with the last but two-layer of contextual transformer block $\mathbb{C}$ of Wav2Vec2.0 with and without prior application of LDA. From the results, we can observe that for SD, the contextual embeddings from Wav2Vec2.0 outperforms in all the classes with an overall relative improvement (TA) of 5.82\% using KNN,	11.54\%	using NBC, 10.38\% using NN over \citep{stutternet} and over~\citep{stutternetmtl} by 11.92\% using KNN, 17.97\% using NBC and 16.74\% using NN. Figure~\ref{fig:layerimpact}. shows the impact of different contextual embeddings in the detection of various stuttering classes. The plot shows almost a close trend in all the stuttering class accuracies. The detection accuracy of stuttering classes increases with the layer\footnote{L1 is a local encoder, L13 is the last layer of $\mathbb{C}$} number. We hypothesize that the lower layers including local encoder (L1) representations contain speech information only from the local window of size 25~ms, and, in addition, passing the representations to the downstream classifiers after applying the statistical pooling layer further restricts it in capturing more stutter specific patterns. In addition, the results show that the contextual layers from L6 to L12 of the Wav2Vec2.0 model trained in a self-supervised fashion are able to capture rich stuttering patterns as also depicted in Fig.~\ref{fig:embed}. As for the last layer (L12), it slightly degrades performance in SD in comparison to its previous layer due to the fact that the transformer block $\mathbb{C}$ was fine-tuned and adapted towards the ASR task. By fine-tuning towards ASR, it is possible that the Wav2Vec2.0 model has not focused on the information which is relevant to stuttering, resulting in the loss of rich stuttering information. Consider such an example of prosodic information, which is very essential in SD, but not that important for ASR. Using Wav2Vec2.0 embeddings with NN in SD, there is an overall relative improvement of 61.36\%, 47.91\%, 14.65\%, and 9.02\% in repetitions, blocks, interjections, and fluents respectively over the MB \emph{StutterNet}~\citep{stutternetmtl}, thus outperforms over the state-of-the-art results. 
Moreover, the prior application of LDA on Wav2Vec2.0 representations further boosts the detection performance in repetitions by 8.29\%, prolongation by 10.41\%, blocks by 27.24\%, and interjections by 8.72\%.

\par 
\textbf{Fusion}: In addition, we fuse the ECAPA-TDNN and Wav2Vec2.0 embeddings via score and embedding fusion schemes, the results of which can be seen in Table~\ref{tab:fusion}. While computing the final score $p$ from ECAPA-TDNN and Wav2Vec2.0 prediction probabilities, we empirically optimize the weighting parameter $\alpha$ on test set in $p = \alpha * p_{w2v2} + (1 - \alpha) * p_{ecapa}$ and we found $\alpha = 0.9$ gives the best results. The ECAPA-TDNN representations which contain rich information about speakers' identity further enhances the overall detection performance of 2.1\% using KNN, 0.65\% using NBC, and 1\% using NN. However, it doesn't contain enough rich information about suprasegmental, emotional content, etc, which are incredibly important for disfluent classes, thus acting as a negative transfer for some of them. Moreover, concatenating the ECAPA-TDNN and Wav2Vec2.0 speech embeddings results in a further relative improvement of 8.66\% in UAR using NN. 
\par 
Each layer of the Wav2Vec2.0 model contains different speech representations, exploiting this fact, we integrate the representations after applying LDA (of component size four) from the local encoder (L1), L7 and L11 from contextual block $\mathbb{C}$, resulting in a $\mathbb{R}^{N\times12}$ dimensional data. 
Concatenating information from multiple layers further improves the minority class recognition including prolongations and blocks by a relative margin of 7.29\% and 29.74\% respectively using NN. Moreover, the embedding fusion also enhances the UAR by a relative margin of 19.17\%, 36.84\%, 46.18\% over the MFCC-based baselines \emph{StutterNet}~\citep{stutternet}, MB \emph{StutterNet}~\citep{stutternetmtl}, and MB ResNet+BiLSTM~\citep{tedd} respectively.


\section{Conclusion}  
The automated stuttering detection task suffers from a lack of unlabeled data and thus limits the application of large deep models. To address this issue, we introduced a self-supervised learning framework that first learns a feature extractor for a pre-text task on a large amount of audio data and then employs the learned feature extractor for the downstream stuttering detection task with limited audio data. We investigated ECAPA-TDNN-based speaker recognition and Wav2Vec2.0-based speech recognition as two separate pre-text tasks trained on VoxCeleb and LibriSpeech, respectively. Our study reveals that contextual embeddings associated with speech recognition tasks are more appropriate for SD than speaker embeddings. We found that Wav2Vec2.0-based contextual embeddings yielded at least 19.17\% relative improvement in UAR over the competitive state-of-the-art systems trained only on limited labeled data. We further improved SD performance by combining embedding from different layers of the Wav2Vec2.0 model. We also found that post-processing the extracted embeddings with LDA improves classification performance. Our benchmarking experiments with three different classifiers for downstream tasks reveal that a simple MLP-based neural network performs best and it opens up opportunities for further advancements.
\par 
Even if the proposed self-supervised framework substantially improved the SD performance over baselines, the performance is still relatively low, possibly due to the presence of a mismatch effect in podcast recordings and data imbalance during training. The future work includes further investigations of the proposed approach by compensating for those effects. In our present work, we did not utilize the stuttering labeled dataset for training the pre-text task. This work can also be extended by including this dataset in this stage which may mitigate the audio domain mismatch between pre-text and downstream tasks. \textcolor{black}{It is possible that the proposed system might fail in the cross lingual setup, and it would also be interesting to analyse the performance of the proposed models in cross lingual setup, where the model is trained on one stuttering language and tested on other. Due to inter-person variances, language/accent/dialect variability, and other speaking variations, stuttering identification is inherently a difficult process. In order to further enhance the stuttering detection systems, speaker-adaptive training and domain adaptation techniques can be exploited to learn these meta-data invariant features.}

\section*{Acknowledgements}
This work was made with the support of the French National Research Agency, in
the framework of the project ANR BENEPHIDIRE (18-CE36-0008-03). Experiments
presented in this paper were carried out using the Grid’5000 testbed, supported by a
scientific interest group hosted by Inria and including CNRS, RENATER and several
universities as well as other organizations(see https://www.grid5000.fr) and using the
EXPLOR centre, hosted by the University of Lorraine.

\section*{Conflict of Interest}

The authors declare that they have no known competing financial interests or personal relationships that could have appeared to influence the work reported in this paper

\section*{Data Availibity}

The dataset analysed during the current study is publicly available from~\citep{sep28k}.

\bibliographystyle{aps-nameyear}      
\bibliography{mybib.bib}

\end{document}